\documentclass[aps,epsfig,graphics, twocolumn]{revtex4}
\usepackage{graphicx,amsmath}

\def\lsim{\lower.5ex\hbox{$\; \buildrel < \over \sim \;$}}
\def\gsim{\lower.5ex\hbox{$\; \buildrel > \over \sim \;$}}
\def\g{\ifmmode \gamma \else $\gamma$\fi}
\def\gs{\ifmmode \gamma \else $\gamma~$\fi}

\newcommand{\la}{\langle}
\newcommand{\ra}{\rangle}

\begin{document}

\title{Canonical suppression in microscopic transport models}

\author{O. Fochler, S.Vogel, M. Bleicher, C. Greiner, P. Koch-Steinheimer, 
Z. Xu}

\affiliation{Institut f\"ur Theoretische Physik, J.W. Goethe Universit\"at \\
Max-von-Laue-Str. 1, D-60438 Frankfurt am Main, Germany}

\begin{abstract}
We demonstrate the occurrence of canonical suppression 
associated with the conservation of an U(1)-charge 
in current transport models. For this study a pion gas is simulated 
within two different transport approaches by incorporating
inelastic and volume-limited collisions $\pi\pi\leftrightarrow
K\overline{K}$ for the production of kaon pairs.
Both
descriptions can dynamically account for the suppression in
the yields of rare strange particles in a limited box, being 
in full accordance with a canonical statistical description.

\end{abstract}

\maketitle

\section{Introduction and Motivation}

Since long, the various properties of excited nuclear matter occurring in heavy collisions
have been studied within statistical descriptions 
\cite{Fermi:1951,Landau:1953gs,Hagedorn:1970gh,Siemens:1979dz,Mekjian:1982xi,Hahn:1986mb}. Here the underlying principle is the stringent 
assumption of full thermal equilibrium in the stage of 
the reaction to which the description is applied. 

In the last 10 years thermal model ansatzes 
of hadronic resonance gases via a grand canonical description have become very popular again because detailed experimental 
results for the yields of individual hadronic species became available
at various bombarding energies. Indeed, it was found that the 
statistical description can be applied very successfully for ultrarelativistic 
energies, indicating that the individual
hadronic particles seem to evolve to a point of nearly perfect chemical
equilibrium \cite{Braun-Munzinger:1995bp,Braun-Munzinger:2001ip}
(for a comprehensive review the reader is referred to \cite{Braun-Munzinger:2003zd}).

However, for low collision energies (e.g. in the SIS, AGS and FAIR range)  
especially the strange particles do become rare.
Thus, as has been well known for a long time, a 
canonical instead of a grand canonical description 
of the strange particles has to be applied. 
The canonical approach results in a suppression in the yields for rare
particles in comparison to the grand canonical description when treating the conservation of the corresponding
U(1)-charge exactly
\cite{Rafelski:1980gk,Koch:1982ij,Hagedorn:1984uy}.
Indeed, thermal models including
the canonical suppression have been applied rather successfully
for the description of the few measured hadronic yields in heavy ion collisions  
at SIS energies \cite{Cleymans:1997sw,Cleymans:1998yb,Oeschler:2000dp}.

A dynamical interpretation of the canonical suppression 
has recently also been offered by the formulation and solution of
kinetic master-equations \cite{Ko:2000vp,Redlich:2001wc}.
It is the purpose of this investigation to show that such
an occurrence of canonical suppression for rare particles
is already warranted by present day transport models.

A priori this is not a trivial statement. One could argue  
that the transport models are based on
solving a set of coupled Boltzmann equations which originate from a grand canonical
treatment. However, this is not the case as the realisations do conserve
energy exactly and also the individual quantum charges like baryon number
and (net) strangeness are conserved during the propagation and scattering processes as well.

In the following section we briefly review the canonical suppression of rare 
particles associated with the conservation of an U(1)-charge and
we also summarize its dynamical formulation via a master equation 
as given in \cite{Ko:2000vp,Redlich:2001wc}. In Section III we
describe our simulation setup. A pion gas is simulated 
by incorporating inelastic and volume-limited collisions $\pi\pi\leftrightarrow
K\overline{K}$ for the production of kaon pairs.
We employ two
completely different numerical realisations for the treatment of  
collisions. The first model under investigation is UrQMD \cite{Bass:1998ca, Bleicher:1999xi}, 
where two-particle collisions are realized via a standard geometrical
interpretation. For previous studies of the thermodynamic properties of hadronic matter within the UrQMD model, the reader is referred to \cite{Belkacem:1998gy, Bravina:1999dh, Bravina:2000dk, Bravina:2000iw}. In the second model the collisions are treated
by transition rates in small spatial subcells. The latter algorithm has most recently been successfully introduced in a covariant parton cascade to
describe inelastic multiparticle  Bremsstrahlung processes of type
$gg \leftrightarrow ggg$ \cite{Xu:2004mz}. In Section IV we present the results of our analysis.
Section V then provides a summary and a conclusion.
We will argue that a (canonical) chemical equilibrium
of kaons can by far not be 
achieved at intermediate
energies for relativistic heavy ion collisions, at least if known
cross sections for the production of kaons are assumed.

\section{Canonical Suppression and its Dynamical Description via a Master 
Equation} \label{secTheory} 

The statistical description of systems incorporating the exact
conservation of quantum numbers has been established for many
years and there exist several approaches of varying complexity and
generality \cite{Rafelski:1980gk,Koch:1982ij,Hagedorn:1984uy}.
Here we give a brief summary of the results
adapted to our needs. The conserved U(1)-charge to be considered 
is strangeness
whose net value is taken to be zero throughout our calculations and
simulations. Explicitly we consider 
only inelastic reactions of the type
$\pi\pi\leftrightarrow K\overline{K}$, where the kaons and anti-kaons bear
strangeness $+1$ and $-1$ respectively.

For a large number of kaons it is sufficient to treat strangeness
conservation on the average $\langle N_{K} \rangle - \langle
N_{\overline{K}} \rangle = 0$ (with $N_{K}$, $N_{\overline{K}}$ being the number
of kaons and anti-kaons, respectively). Introducing a chemical potential $\mu_{s}$,
or fugacity $\lambda_{s}=\exp\left(\frac{\mu_{s}}{T}\right)$, to
control the net strangeness content, one is led to the following 
grand canonical partition function
\begin{equation} \label{gcZ}
Z^{\mathrm{gc}}\left(V,T,\lambda_{s}\right)=\exp\left(Z^{1}_{\pi}+\lambda_{s}Z^{1}_{K}+\lambda^{-1}_{s}Z^{1}_{\overline{K}}\right), 
\end{equation}  
where $Z^{1}_{i}$ denotes the relativistic one-particle partition
function for non-interacting particles of type $i$ (pions or kaons in this case), $V$ is the volume, $T$ is the temperature and $m_i$ is the particle's mass. $K_2$ denotes a modifed Bessel function. 
\begin{equation} \label{1partZ}
Z^{1}_{i}\left(V,T\right)=g_{i}\frac{VT}{2\pi^{2}}m_{i}^{2}K_{2}\left(\frac{m_{i}}{T}\right).
\end{equation}
In order to consider a more general case, the one-body partition
functions in (\ref{gcZ}) would have to be replaced by sums over
one-body partition functions of all hadronic particles carrying the
corresponding integer in strangeness of $0$, $\pm 1$, $\pm 2$ and
$\pm 3$.

A small number of kaons (roughly spoken when the average number
becomes of order one) demands the conservation 
of strangeness
to be treated exactly, i.e. $N_{K}-N_{\overline{K}}=0$ for
each state contributing to the partition sum. This constraint leads to a
reduction of the available phase-space for the production process and
ultimately one obtains the canonical partition function describing the
system via ($I_0$ and $I_1$ denote Bessel functions)
\begin{equation} \label{cZ}
Z^{\mathrm{c}}\left(V,T\right)=\exp\left(Z^{1}_{\pi}\right)I_{0}\left(x\right)
\end{equation}
with
\begin{equation} \label{x}
x=2\sqrt{Z^{1}_{K}Z^{1}_{\overline{K}}}=2Z^{1}_{K}.
\end{equation}

From  (\ref{gcZ}) and  (\ref{cZ}) one can calculate the density $n_K$ of particles
with strangeness $+1$, the kaons, via the relation:
\begin{equation} \label{gcdens}
n^{\mathrm{gc}}_{K}=\left.\frac{1}{V}\lambda_{s}\frac{\partial}{\partial\lambda_{s}}\ln
Z^{\mathrm{gc}} \right\vert_{\lambda_{s}=1} 
= \frac{1}{V}\lambda_{s}Z_{K}^{1}=\frac{Z_{K}^{1}}{V}
\end{equation}
and considering the canonical case one finds 
\begin{align} \label{cdens}
n^{\mathrm{c}}_{K}&=\left.\frac{1}{V}\lambda_{s}\frac{\partial}{\partial\lambda_{s}}\ln
Z^{\mathrm{c}}\right\vert_{\lambda_{s}=1}\nonumber\\
&= \frac{1}{V}\frac{I_{1}\left(x\right)}{I_{0}\left(x\right)}\sqrt{\frac{Z_{\overline{K}}^{1}}{Z_{K}^{1}}}Z_{K}^{1} = \eta\ n_{K}^{gc},
\end{align}
using that in the grand canonical picture $\langle N_{K} \rangle - \langle
N_{\overline{K}} \rangle = 0$ requires $\lambda_{s}=\sqrt{Z_{\overline{K}}^{1}/Z_{K}^{1}}=1$. 

Comparing (\ref{gcdens}) and (\ref{cdens}) one defines the canonical suppression factor $0\leq\eta\leq1$
\begin{equation} \label{eta}
\eta=\frac{n_{K}^{\mathrm{c}}}{n_{K}^{\mathrm{gc}}}=\frac{I_{1}(x)}{I_{0}(x)},
\end{equation}
which contains all relevant information on differences in the equilibrium 
particle
density between the grand canonical and canonical description.

An alternative way of understanding this suppression is to consider kinetic master equations
\cite{Ko:2000vp,Redlich:2001wc} by
looking at a single process $ab\leftrightarrow c\bar{c}$. $P_{N_{c}}(t)$
denotes the probability to find $N_{c}$ particles $c$ at a time
$t$. In our case we have $c \equiv K$ and thus 
$N_{c}=N_{\bar{c}}$ holds exactly, whereas particles
$a$ and $b$ (the `pions') are assumed to be uncorrelated and abundant. Furthermore, the
probabilities for a single gain or loss process per unit time and volume are
denoted by $G/V$ and $L/V$ respectively, where
$G=\langle\sigma_{G}v\rangle$ and $L=\langle\sigma_{L}v\rangle$ are
the momentum-averaged cross sections for the gain and loss
processes. With that, a master equation can be formulated \cite{Ko:2000vp}
\begin{align} \label{mastereq}
\frac{\mathrm{d}P_{N_{c}}}{\mathrm{d}t}&=\frac{G}{V}\la N_{a}\ra\la
N_{b}\ra P_{N_{c}-1}+\frac{L}{V}\left(N_{c}+1\right)^{2}P_{N_{c}+1}\nonumber\\
&-\frac{G}{V}\la N_{a}\ra\la
N_{b}\ra P_{N_{c}}+\frac{L}{V}N_{c}^{2}P_{N_{c}}.
\end{align}

A general solution of this equation is possible \cite{Redlich:2001wc}, but for our purpose it
is sufficient to look at the kinetic equation for the time evolution
of the average number $\la N_{c}\ra$. It can be obtained when
multiplying (\ref{mastereq}) by $N_{c}$ and summing over it:
\begin{equation}\label{kineq}
\frac{\mathrm{d}\la N_{c}\ra}{\mathrm{d}t}=\frac{G}{V}\la N_{a}\ra\la
N_{b}\ra-\frac{L}{V}\la N_{c}^{2}\ra.
\end{equation}
This equation can be easily treated in the two limiting cases $\la
N_{c}\ra\gg 1$ and $\la N_{c}\ra\ll 1$. For abundant production of
$c\bar{c}$ pairs, i.e.\ when  $\la N_{c}\ra\gg 1$, the relation $\la
N_{c}^{2}\ra\approx\la N_{c}\ra^{2}$ holds. 
On the other hand, for very rare production, i.e.\ when $\la
N_{c}\ra\ll 1$, one has the relation $\la
N_{c}^{2}\ra\approx\la N_{c}\ra $. 
Assuming a thermal momentum distribution with
\begin{equation}
\frac{G}{L}=\frac{Z^{1}_{c}Z^{1}_{\bar{c}}}{Z^{1}_{a}Z^{1}_{b}},
\end{equation}
and looking at stationary solutions for large times (the solutions
describing an equilibrated system) one ends
up with ($\epsilon\equiv G\la N_{a}\ra\la N_{b}\ra/L$)
\begin{equation}\label{kinsolgc}
n_{c}= \frac{ \la N_c \ra }{V} = 
\frac{\sqrt{\epsilon}}{V}=\frac{Z^{1}_{c}}{V} \equiv n_c^{gc},  
\end{equation}
for $\la N_{c}\ra\gg 1$. The opposite case $\la N_{c}\ra\ll 1$ leads
to
\begin{equation}\label{kinsolc}
n_{c}= \frac{ \la N_c \ra }{V} = 
\frac{\epsilon}{V}=\frac{Z^{1}_{c}Z^{1}_{\bar{c}}}{V} \equiv n_c^{c}  .
\end{equation}

Identifying particle-type $c$ with kaons, it is clear that
(\ref{kinsolgc}) equals the grand canonical result
(\ref{gcdens}). (\ref{kinsolc}) is just the leading term in an
expansion of the canonical result (\ref{cdens}). Hence, it is
verified that an abundant production of particles leads to a
grand canonical description, whereas rare particle production requires
a canonical description.

Let us have a closer look at the canonical suppression factor (\ref{eta})
using the asymptotic behaviours
\begin{align}\label{asymptI}
\lim_{x\to\infty}\frac{I_{1}(x)}{I_{0}(x)}&\rightarrow 1&
\lim_{x\to 0}\frac{I_{1}(x)}{I_{0}(x)}&\rightarrow \frac{x}{2} .
\end{align}

\begin{figure}[hbt]
\centering
\includegraphics[width=.5\textwidth,angle=0]{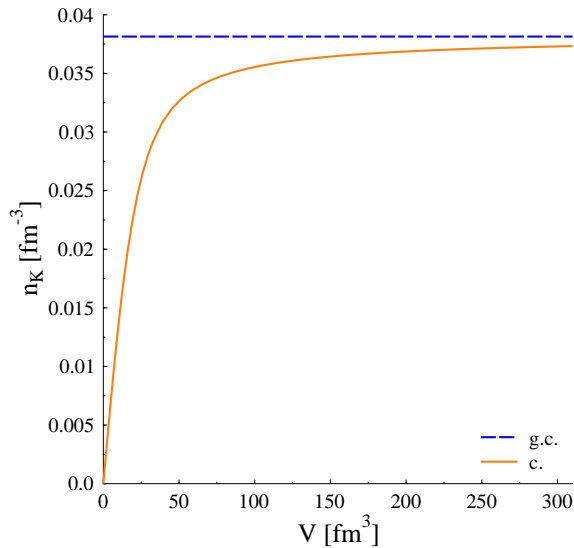}
\caption{Volume dependence of the kaon density 
for a canonical ensemble (cf. eq. (\protect{\ref{cdens}})) 
for $T=170\ \mathrm{MeV}$. g.c. denotes the grand canonical calculation, c. labels the canonical result.}
\label{fig_densTheory}
\end{figure}

Here, one sees that in the grand canonical limit the kaon density $n_{K}$ is independent
of the reaction volume, whereas in the canonical regime with the number of kaons $\la N_{K}\ra\ll 1$,
it scales linearly with the volume as $x\propto
V$. Fig.~\ref{fig_densTheory} illustrates this behaviour.  

\section{Real Transport and Simulation Setup}

The main objective of our study is to demonstrate that current 
(large scale) transport models are able to reproduce the effect
of canonical suppression. This is not a trivial investigation. One might argue  
that the transport algorithms are based on
strictly solving the Boltzmann equation which in turn originates from a grand canonical
treatment. In the stationary limit 
of the (coupled) Boltzmann equations the equilibrium phase space distributions,
where the various collision terms 
for each particle population vanish, obey Maxwell-Boltzmann 
statistics (or Bose-Einstein
or Fermi-Dirac statistics when the Pauli factors are included), so that
the individual densities are that of a grand canonical ensemble. 
The reason for this is that for describing an annihilation 
within a Boltzmann description the two (rare) particles are assumed
to be completely uncorrelated. This is however not the case in the description
via the master equation (\ref{mastereq}). Here the occurrence 
of a pair of rare particles is explicitly taken care of. Due to the nature of pair production
the existence of a kaon requires the existence of an anti-kaon. This leads
to an effective increase in the annihilation process compared
to the Boltzmann  Stosszahlansatz. Inspecting the loss
term in (\ref{kineq}), the approximation leading to a Boltzmann process
is given by
\begin{equation}
\label{Boltzappr}
\la
N_{c}^{2}\ra \, \stackrel{\mbox{Boltzmann process}}{\rightarrow }
\la N_{c}\ra^{2} \, ,
\end{equation}
which then leads to the grand canonical description.
Staying within the description of the master equation,
the possibility for annihilation is much higher. When
there exists a kaon, there has to be an anti-kaon with which
an annihilation process is possible. $\la N_c^2 \ra \gg \la N_{c}\ra^{2} $ for
$\la N_c \ra \ll 1$, hence the effective annihilation rate 
incorporated correctly in the master equation is much larger than
for a Boltzmann description, leading to the canonical suppression.  

On the other hand the various numerical realisations 
of the underlying transport equations do conserve energy and the individual quantum charges (like baryon number
and net strangeness) exactly - as within a microcanonical 
treatment. The argument of an `enhanced' annihilation should
therefore also apply within these realisations. In order to address this question, we
will concentrate on the volume dependencies in the grand canonical and canonical regime respectively, as depicted in Fig. \ref{fig_densTheory}.

\begin{figure}[hbt]
\centering
\includegraphics[width=.5\textwidth,angle=0]{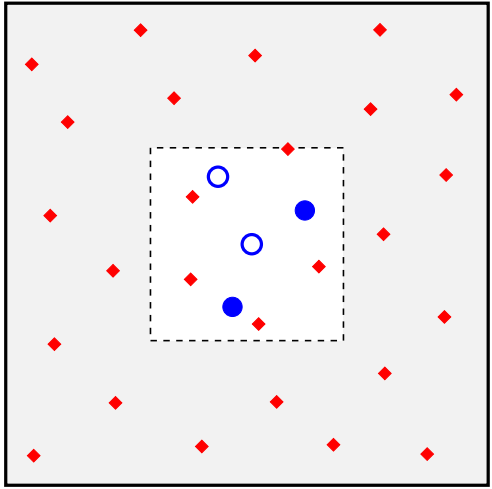}
\caption{Illustration of the two boxes for the numerical simulation:
The pions (diamonds)  move inside the larger and fixed box and thus provide a heat bath for the kaons and anti-kaons (open and filled circles),
which can only move and interact inside the small box. 
The size of the small box is varied for investigating
the dynamical occurrence of canonical suppression in small
volumes. }
\label{fig_illustration}
\end{figure}

The simulation setup consists of a large box of $20\ \mathrm{fm}$
side length holding a relativistic gas of pions. The pion gas provides
a heat bath for a much smaller reaction volume of variable size centered within
the large box. Inside this smaller and likewise box-shaped reaction volume
processes 
$$\pi\pi\leftrightarrow K\overline{K}$$
are allowed, covering all
possible isospin states of pions and kaons. The kaons are
reflected by the walls of the small reaction volume and are thus bound to
it, whereas these walls are permeable for the pions. 
An illustration of this special spatial setting is given in Figure
\ref{fig_illustration}.
After equilibration, the kaon density within the reaction volume should be
governed by eq. (\ref{cdens}) which holds as a reference for the transport
model results. The idea therefore is to simulate different sizes of the inner reaction volume and to extract the number of
kaons $\la N_{K}\ra$ for each simulation run by averaging over many
timesteps in order to minimise the statistical fluctuations. These
timesteps are sufficiently separated to avoid correlations. Furthermore the data are taken after the equilibration of
the system.
 
We implemented this scenario using two different types of transport
descriptions - the microscopic transport model UrQMD
\cite{Bass:1998ca,Bleicher:1999xi} and a realisation of a stochastic
transport model borrowed from a recently developed parton cascade
\cite{Xu:2004mz}.
The former model makes use of a geometrical interpretation of cross sections in order to solve the
transport equations, whereas the latter relies on the explicit calculation of
transition probabilities. 

For standard applications, the UrQMD model provides full space time dynamics for hadrons and
strings. It is a non-equilibrium model based on the covariant
propagation of hadrons and strings. All cross sections are fitted 
to available data or calculated by the principle of
detailed balance. For our studies the code is modified such that only
reactions $\pi\pi\leftrightarrow K\overline{K}$, together with elastic
collisions among the pions and kaons, remain possible. 
The $\pi\pi\rightarrow K\overline{K}$
inelastic reactions are assigned a constant cross section of either $1
\mathrm{mb}$, $5 \mathrm{mb}$ or $10 \mathrm{mb}$, whereas the
backward reactions are calculated via the principle of detailed balance. 

The simulation of $2\leftrightarrow 2$ processes within the stochastic
method is based on the calculation of a collision probability for each
possible particle pair per unit volume $\Delta ^3 x$ and unit time 
$\Delta t$ via \cite{Xu:2004mz}
\begin{equation} \label{collProb}
P_{22}=v_{rel}\sigma_{22}\frac{\Delta t}{\Delta x^{3}}.
\end{equation}
Here, $v_{rel}$ denotes the relative velocity and $\sigma_{22}$ is the cross
section for the considered $2\leftrightarrow 2$ process. Similar to the
UrQMD setup, the cross sections are set to be constant in one
direction and calculated via detailed balance for the reverse
reaction.
The so obtained probability is then compared with a random number
between $0$ and $1$ to decide whether the collision does take place
or not. The implementation of the stochastic model is
therefore closely related to the formulation of the master equation
(\ref{mastereq}) and its solution discussed in section~\ref{secTheory}.
We have to stress one important point: The stochastic method is, in
principle, flexible to introduce test particles, that is to `subdivide'
each particle into a number N of testparticles \cite{Xu:2004mz}.
However, for the following investigation it is crucial that the produced kaons are {\em not}
subdivided into further testparticles.
We will discuss this further below.

The initial conditions in both schemes are chosen such that the pion
gas acquires a temperature of $T=170\ \mathrm{MeV}$. The appropriate
number of pions and the total energy of the system are calculated via
the use of a grand canonical partition function (\ref{gcZ}) for pions
alone, as kaons are absent in the initial state. The total energy
evaluates to
\begin{equation} \label{totalE}
U(T,V)=\frac{g_{\pi}V}{2\pi^{2}}m^{2}_{\pi}T^{2}\left\{
3K_{2}\left(\frac{m_{\pi}}{T}\right) +\frac{m_{\pi}}{T}K_{1}\left(\frac{m_{\pi}}{T}\right)\right\}.
\end{equation} 
A heat bath volume of $8000\ \mathrm{fm}^{3}$, as used in our simulations, 
then 
corresponds to a population of $1348$ pions bearing a total energy 
of $747.5\ \mathrm{GeV}$. Initially each pion is assigned the same
fraction of the total energy, giving one half of the particles momenta
in the positive $x$-direction, while the remaining particles start out
bearing momenta in the negative $x$-direction. The spatial
distribution is random.

\section{Results for the Dynamical Suppression}

\begin{figure} [hbt]
\centering
\includegraphics[width=.5\textwidth,angle=0]{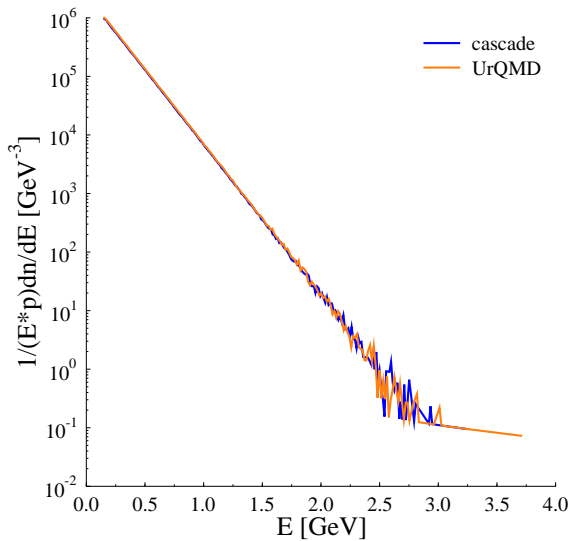}
\caption{Logarithmic energy spectra for the pion gases within the
UrQMD- and cascade-calculations. Both spectra are identical within statistical fluctuations.}
\label{fig_spectrum}
\end{figure}

A necessary verification of the simulations
reliability is to check for
kinetic equilibration. Fig.~\ref{fig_spectrum} demonstrates that the
energy distribution of the pions does become exponential
over six orders of magnitude with the desired
temperature of $T=170\ \mathrm{MeV}$ in both schemes.

\begin{figure} [hbt]
\centering
\includegraphics[width=.5\textwidth,angle=0]{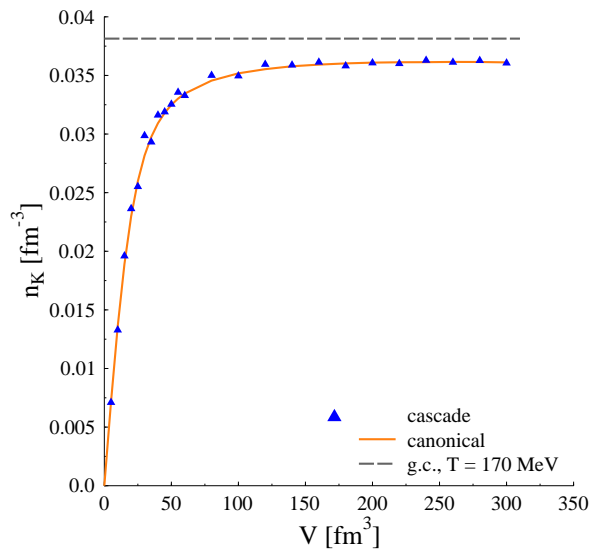}
\caption{Kaon density versus reaction volume as extracted from
simulations within the stochastic cascade (triangles). For comparison, the dashed line indicates the grand
canonical behaviour. The solid line shows the canonical volume dependence of the kaon
density, based on (\ref{cdens}) for a temperature of $T = 170\
\mathrm{MeV}$ and corrected by taking small deviations in temperature
and fugacity into account.}
\label{fig_densSimCascade}
\end{figure}

\begin{figure} [hbt]
\centering
\includegraphics[width=.5\textwidth,angle=0]{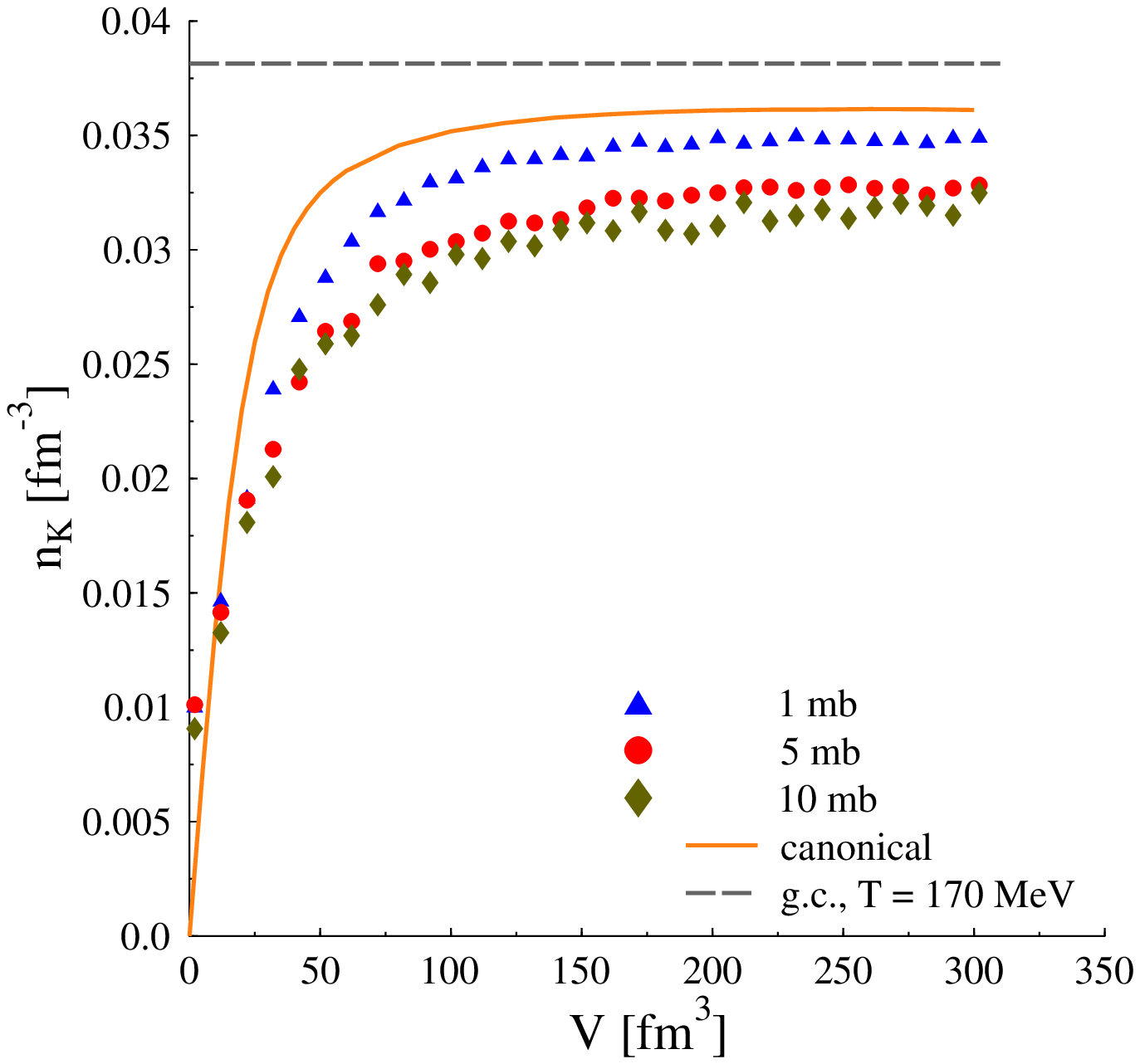}
\caption{Kaon density versus reaction volume as extracted from
simulations using the UrQMD model. Cross sections of $1 \mathrm{mb}$ (triangles), $5
\mathrm{mb}$ (circles) and $10 \mathrm{mb}$ (diamonds) have been assigned to the forward
direction of the inelastic reactions $\pi\pi\leftrightarrow
K\overline{K}$.
The solid and dashed lines again provide the theoretically expected
behaviour as seen in Fig.~\ref{fig_densSimCascade}.}
\label{fig_densSimURQMD}
\end{figure}

Figures~\ref{fig_densSimCascade} and~\ref{fig_densSimURQMD} now 
display the actual results in terms of
the kaon density as a function of the (small)  reaction volume. The minor
fluctuations in the results indicate the
small statistical errors due to the finite number
of averages done. It is clear that the
canonical suppression is reproduced by both transport schemes. 
The kaon
yield is suppressed for small reaction volumes with respect to the
grand canonical limit.  
This verification states the main result of the present study.  

In general, certain deviations from the theoretical values are to be
expected due to the choice of initial conditions. Pion number and
energy content are fixed such that, without any kaons present, the
system is tuned to the reference temperature of $T=170\ \mathrm{MeV}$
and fugacity $\lambda=1$. The production of the heavier kaons then leads to small changes in these parameters, as the total energy 
in the system is conserved as well as the total number of particles,
because only particle number conserving collision processes are included
in the present study.
This slightly affects the
equilibrium number of pions and thus also the kaons. 
For the case of the stochastic cascade we
investigated the relative deviations of temperature
$T_{new}= T + \Delta T $ and effective mesonic fugacity 
$\lambda _{eff} = 1 + \Delta \lambda $ more
closely. It is found that both deviations indeed grow with the number of kaons
produced and thus with the size of the reaction volume. The observed deviations
are on the order of two percent at most.    
Taking these deviations into account, the theoretical results can be
adapted to the actual conditions present in the system at a given size
of the reaction volume 
by rescaling with the effective fugacity factor. 
The actual theoretical reference curves in
Figs~\ref{fig_densSimCascade} and~\ref{fig_densSimURQMD} are
corrected that way.

The stronger deviations for the microscopical UrQMD model cannot be solely
explained by the changes in temperature and fugacity. In fact the
crucial effect when using the geometrical concept of incorporating
binary collisions is a decrease in the collision rate, when the expected
mean free path $\lambda_{\mathrm{m.f.p.}}=(n\sigma_{22})^{-1}$ 
for the particular reaction gets in
the order of the interaction length $\sqrt{\sigma_{22}/\pi}$ as 
pointed out in \cite{Xu:2004mz}. This is due to the difference in the
collision times of the involved particles viewed from the computational
frame. During that interval the particle having the larger collision
time must not collide again to ensure causality. Thus, the collision
rate is decreased compared to the one given by the collision integral. As
different densities are involved, one can not expect forward and
reverse reactions to be affected in the same way and the changes in
reaction rates lead to a shifted stationary `equilibrium' value of the kaon
density.

This behaviour can be clearly seen in Fig.~\ref{fig_densSimURQMD}, where the
kaon density is depicted for different cross sections of the forward reaction
$\pi\pi\rightarrow K\overline{K}$. The smaller the chosen cross section and
thus the larger the mean free path, the more accurately the
theoretical result is reproduced. On the contrary, for typical real
mesonic cross section the numerical shortcoming is in the range 
of 10 to 15 percent.
Too few kaons relative to the pions are simulated in comparison to
the theoretical limit either in the canonical or grand canonical regime.
Such a discrepancy in kaon number relative to the pion
number with respect to experimental results has been reported in 
several URQMD calculations \cite{Bratkovskaya:2004kv}. It might well be that
a better implementation of the collision criteria will enhance the kaon 
relative to the pion yield. We leave this for a detailed future 
investigation.

Another instructive way to analyse the results is to look at the
suppression factor $\eta$ (\ref{eta}) as a function of the average number of
kaons $\la N_{K}\ra^{\mathrm{c}}$ present in the system. This relation
is independent of the temperature which can be easily shown:
Since the suppression factor is a function of $Z^{1}_{K}$,
$\eta=\eta(Z^{1}_{K})$, the one-body partion function can in
principle be expressed in terms of $\eta$,
\begin{equation} \label{ZofEta}
Z^{1}_{K}=Z^{1}_{K}(\eta),
\end{equation}
which then leads to 
\begin{equation} \label{NofEta}
\la N_{K}\ra^{\mathrm{c}}=\eta\la N_{K}\ra^{\mathrm{gc}}=\eta Z^{1}_{K}(\eta).
\end{equation}

\begin{figure} [hbt]
\centering
\includegraphics[width=.5\textwidth,angle=0]{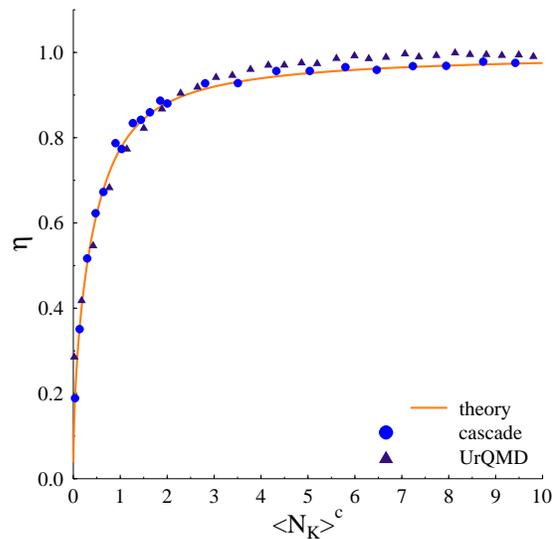}
\caption{Suppression factor $\eta$ as a function of the average kaon number.}
\label{fig_nVsEta}
\end{figure}

Fig.~\ref{fig_nVsEta} states the summary of the investigation and 
depicts  the above functional relation together with
the simulation results. 
Here the temperature and fugacity deviations
have been taken into account for the cascade results. 
For the calculations employing the UrQMD model the suppression factor
has been calculated by taking the ratio of the actual kaon density
$n_K(V)$ divided by the density $n_K(V\rightarrow \infty )$ in the limit 
of very large volumes. Within this prescription
the sensitivity on the cross section vanishes almost completely. 
Apart from the demonstration of
the excellent agreement between theory and simulation, the plot illustrates the
two limiting cases considered when solving the kinetic equation
(\ref{kineq}). Small particle numbers $\la N \ra\ll 1$ are associated
with the canonical regime, whereas for abundant particle production $\la N \ra\gg 1$
the grand canonical description is fully valid.

\section{Summary and Conclusions}

We have demonstrated that two current 
(large scale) transport models are able to reproduce the effect
of canonical suppression. As a particular example we have chosen the production of kaons
and anti-kaons via pions. The overall kaon yield is suppressed with respect to
the grand canonical limit for small reaction volumes and thus small
numbers of produced kaons.

The transport algorithms are based on
solving the Boltzmann equation. Still they 
do conserve energy and net strangeness exactly.
The canonical suppression for kaon numbers $\la N_{K}\ra$ considerably smaller than one, as pointed out in \cite{Ko:2000vp},
originates dynamically from an enhancement of the annihilation process
by $1/\la N_{K}\ra$ as compared to standard, grand canonical
formulation of the Boltzmann equation. The reason is that any kaon in
the system requires the existence of a corresponding
anti-kaon due to strangeness conservation. Thus the probability of
finding a particle anti-particle pair turns into a 
highly correlated conditional
probability compared
to that obtained via the uncorrelated Boltzmann Stosszahlansatz.
The so enhanced annihilation probability then leads to
the canonical suppression in the kaon yields. As the calculations show,
it is manifest that the occurrence of canonical suppression is 
dynamically reproduced by the two
presented schemes for solving the kinetic transport
equations. A small discrepancy, i.e.\ 
a slightly lower kaon yield on the order of 10 to 15 percent, 
is present when employing typical cross sections.
This discrepancy can be traced back to the numerical
realisation of the occurrence of collisions via the standard geometrical
description.
As already emphasized, detailed investigations
within the UrQMD model are in order to clarify how
important such effects are for the understanding
of kaon yields in relativistic heavy ion collisions, as typically too few kaons are produced 
relative to the pions, when compared to experimental results \cite{Bratkovskaya:2004kv}.

Addressing especially the dynamical generation of canonical suppression,
one has to worry about even more significant violations 
when other numerical transport schemes are invoked:
One popular scheme is the so called test-particle method,
i.e. each real particle is subdivided into $N$ test-particles
in order to obtain smoother distributions or to suppress other
numerical artefacts (see e.g. \cite{Xu:2004mz}). When applying
this idea of particle subdivision to the stochastic method, it will 
rescale the volumes V of Figure \ref{fig_densSimCascade}
by a factor $1/N$, so that the grand canonical limit is achieved
for much smaller volumes. This is, of course, unphysical.
Thus, whenever the canonical suppression is of relevance
for the understanding of particular yields, 
the test-particle method has to be avoided.

Another oftenly applied method is the so called perturbative method
with `virtual' particles (for a discussion see e.g. \cite{Hartnack:2005eh}).
Here it is assumed that the exotic particles being produced are
so rare that their behaviour does not alter the overall dynamics.
If the backreaction of kaon production, i.e. the annihilation
of a kaon and its antiparticle, is not considered correctly,
then detailed balance is violated and the system can never achieve
full and correct chemical equilibrium.

When simulating real heavy ion collisions to address
the production of kaons for lower or moderate relativistic energies
(for a very recent work see \cite{Larionov:2005kz}),
one has indeed to worry whether these backreactions are
important in the sense that the rates to achieve chemical equilibrium
are smaller or comparable to the overall lifetime
of the fireball. Various applications of thermal models 
for such low energies have been put
forward in several works over the last years
for describing hadronic yields in heavy ion collisions  
at SIS \cite{Cleymans:1997sw,Cleymans:1998yb,Oeschler:2000dp,Averbeck:2000sn}.
As only very few yields are available, the soundness of such
an analysis is not evident. In principle, such an analysis rests
on the assumption that chemical equilibrium among the 
individual hadronic particles is temporarily achieved, which
implies that the corresponding time scales are sufficiently short
compared to the lifetime of the fireball.

From transport theory residing on binary elastic and inelastic 
hadronic collisions
it has been known for a long time that for intermediate to moderate energies
the kaons are predominantly produced 
before the initial motion is substantially degraded and
when the system is still far from any
(quasi-)equilibrium stage (for a discussion and review see e.g. 
\cite{Greiner:2001uh}). Putting it differently, when the
momenta of the nucleons are sufficiently degraded and the system
has to some extent thermalized, the timescale for production of
strange particles via the considered inelastic kinetic reactions 
becomes extremely
large. For a particular calculation in a static environment at somewhat
larger energies it was found that the equilibration  
of the kaons is exceeding the lifetime of a potential fireball
by at least two orders of magnitude
\cite{Bratkovskaya:2000qy}. In this analysis the comparison
was made, however, to a grand canonical estimate.

On the other hand, as we will argue now, 
a chemical equilibrium situation of kaons with canonical suppression included
cannot be 
achieved even temporarily when assuming known
cross sections. Within a grand canonical environment the
rate for chemical equilibration of a particle C for a reaction of type
$A+B \leftrightarrow C+D$ is given by $\tau ^{-1} \approx \la \sigma _{CD}
v_{CD} \ra n_D $, where $n_D$ is the particle density of species $D$.
With $n_K = n_{\overline{K}}$
if there is only one kaon, then there also has to exist one anti-kaon in the 
particular realisation of the system. Hence, for such a situation of
a canonical realisation, where the reactions $ \pi + \pi \leftrightarrow K + \overline{K}$ will drive the kaons to chemical equilibrium, the subsequent
equilibration rate is obtained by substituting for $n_D$
the effective anti-kaon density $1/V$. The rate thus reads
\begin{equation}
\label{chemrate_can}
\tau ^{-1} _{\rm{chem}; K} \, \approx \,  
\la \sigma _{K\overline{K}\rightarrow \pi \pi }
v_{K\overline{K}} \ra \frac{1}{V} \, \, .
\end{equation}
In an UrQMD simulation the typical averaged cross section rather close
to threshold is given by 
$\la \sigma _{K\overline{K}\rightarrow \pi \pi }
v_{K\overline{K}} \ra \approx 2 \rm{mb} $, 
so that 
$\tau  _{\rm{chem}; K} \approx 5 V/(\rm{fm}^2 c)$. 
Any typical reaction volume at intermediate time scales
in a relativistic heavy ion reaction is in the order of 100 $\rm {fm}^3$.
Hence, kaons can not be expected to come or stay close to chemical equilibrium.

\section{acknowledgements}

The authors thank S.~Leupold and U.~Mosel for discussions, which
have stimulated this work. This work is supported by GSI, DFG and BMBF. The computational resources have been provided by the Center for Scientific Computing in Frankfurt.



\begin{thebibliography}{00}


\bibitem{Fermi:1951}
E.~Fermi,
Prog.\ Theor.\ Phys. {\bf 5} (1950) 570,
Phys.\ Rev.\ {\bf 81} (1951) 683


\bibitem{Landau:1953gs}
L.~D.~Landau,
Izv.\ Akad.\ Nauk Ser.\ Fiz.\  {\bf 17} (1953) 51.


\bibitem{Hagedorn:1970gh}
R.~Hagedorn,
Nucl.\ Phys.\ B {\bf 24} (1970) 93.


\bibitem{Siemens:1979dz}
P.~J.~Siemens and J.~I.~Kapusta,
Phys.\ Rev.\ Lett.\  {\bf 43} (1979) 1486.


\bibitem{Mekjian:1982xi}
A.~Z.~Mekjian,
Nucl.\ Phys.\ A {\bf 384} (1982) 492.

\bibitem{Hahn:1986mb}
  D.~Hahn and H.~Stoecker,
  Nucl.\ Phys.\ A {\bf 476}, 718 (1988).



\bibitem{Braun-Munzinger:1995bp}
P.~Braun-Munzinger, J.~Stachel, J.~P.~Wessels and N.~Xu,
Phys.\ Lett.\ B {\bf 365} (1996).

\bibitem{Braun-Munzinger:2001ip}
  P.~Braun-Munzinger, D.~Magestro, K.~Redlich and J.~Stachel,
  Phys.\ Lett.\ B {\bf 518}, 41 (2001)

\bibitem{Braun-Munzinger:2003zd}
  P.~Braun-Munzinger, K.~Redlich and J.~Stachel,
  arXiv:nucl-th/0304013.



\bibitem{Rafelski:1980gk}
J.~Rafelski and M.~Danos,
Phys.\ Lett.\ B {\bf 97} (1980) 279.

\bibitem{Koch:1982ij}
  P.~Koch, J.~Rafelski and W.~Greiner,
  Phys.\ Lett.\ B {\bf 123}, 151 (1983).


\bibitem{Hagedorn:1984uy}
R.~Hagedorn and K.~Redlich,
Z.\ Phys.\ C {\bf 27}, 541 (1985).



\bibitem{Cleymans:1997sw}
  J.~Cleymans, D.~Elliott, A.~Keranen and E.~Suhonen,
  Phys.\ Rev.\ C {\bf 57}, 3319 (1998)
  [arXiv:nucl-th/9711066].

\bibitem{Cleymans:1998yb}
  J.~Cleymans, H.~Oeschler and K.~Redlich,
  Phys.\ Rev.\ C {\bf 59}, 1663 (1999).


\bibitem{Oeschler:2000dp}
H.~Oeschler,
J.\ Phys.\ G {\bf 27}, 257 (2001).



\bibitem{Ko:2000vp}
C.~M.~Ko, V.~Koch, Z.~W.~Lin, K.~Redlich, M.~A.~Stephanov and X.~N.~Wang,
Phys.\ Rev.\ Lett.\  {\bf 86}, 5438 (2001).


\bibitem{Redlich:2001wc}
K.~Redlich, V.~Koch and A.~Tounsi,
Nucl.\ Phys.\ A {\bf 702}, 326 (2002).


\bibitem{Bass:1998ca}
S.~A.~Bass {\it et al.},
Prog.\ Part.\ Nucl.\ Phys.\  {\bf 41} (1998) 225.


\bibitem{Bleicher:1999xi}
M.~Bleicher {\it et al.},
J.\ Phys.\ G {\bf 25} (1999) 1859.


\bibitem{Belkacem:1998gy}
  M.~Belkacem {\it et al.},
  Phys.\ Rev.\ C {\bf 58} (1998) 1727
  [arXiv:nucl-th/9804058].
  
  

\bibitem{Bravina:1999dh}
  L.~V.~Bravina {\it et al.},
  Phys.\ Rev.\ C {\bf 60} (1999) 024904
  [arXiv:hep-ph/9906548].
  
\bibitem{Bravina:2000dk}
  L.~V.~Bravina {\it et al.},
  Phys.\ Rev.\ C {\bf 63} (2001) 064902
  [arXiv:hep-ph/0010172].
  
\bibitem{Bravina:2000iw}
  L.~V.~Bravina {\it et al.},
  Phys.\ Rev.\ C {\bf 62} (2000) 064906
  [arXiv:nucl-th/0011011].


\bibitem{Xu:2004mz}
  Z.~Xu and C.~Greiner,
  Phys.\ Rev.\ C {\bf 71}, 064901 (2005).


\bibitem{Bratkovskaya:2004kv}
  E.~L.~Bratkovskaya {\it et al.},
  Phys.\ Rev.\ C {\bf 69} (2004) 054907
  [arXiv:nucl-th/0402026].
  
  

\bibitem{Hartnack:2005eh}
C.~Hartnack,
arXiv:nucl-th/0507002.

\bibitem{Larionov:2005kz}
  A.~B.~Larionov and U.~Mosel,
  Phys.\ Rev.\ C {\bf 72}, 014901 (2005).


\bibitem{Averbeck:2000sn}
  R.~Averbeck, R.~Holzmann, V.~Metag and R.~S.~Simon,
  Phys.\ Rev.\ C {\bf 67}, 024903 (2003).

\bibitem{Greiner:2001uh}
  C.~Greiner,
  J.\ Phys.\ G {\bf 28}, 1631 (2002).


\bibitem{Bratkovskaya:2000qy}
  E.~L.~Bratkovskaya, W.~Cassing, C.~Greiner, M.~Effenberger, U.~Mosel 
  and A.~Sibirtsev,
  Nucl.\ Phys.\ A {\bf 675}, 661 (2000).





\end{thebibliography}
\end{document}